\documentclass[english]{article}
\usepackage[T1]{fontenc}
\usepackage[latin9]{inputenc}
\usepackage{amsbsy}
\usepackage{esint}

\makeatletter

\newcommand{\noun}[1]{\textsc{#1}}

\makeatother

\usepackage{babel}
\begin{document}

\title{Inequalities and Approximations for Fisher Information in the Presence
of Nuisance Parameters}

\author{Eric Clarkson}
\maketitle
\begin{abstract}
Many imaging systems are used to estimate a vector of parameters associated
with the object being imaged. In many cases there are other parameters
in the model for the imaging data that are not of interest for the
task at hand. We refer to these as nuisance parameters and use them
to form the components of the nuisance parameter vector. If we have
a prior probability distribution function (PDF) for the nuisance parameter
vector, then we may mariginalize over the nuisance parameters to produce
a conditional PDF for the data that only depends on the parameters
of interest. We will examine this approach to develop inequalities
and approximations for the FIM when the data is affected by nuisance
parameters.
\end{abstract}

\section{Introduction}

The Fisher Information Matrix (FIM) is important for computing figures
of merit for imaging systems on specific tasks. For estimation tasks
the Cramer Rao Bound (CRB), which is derived from the inverse of the
FIM, provides a lower bound for the covariance matrix of an unbiased
estimator of the parameter vector of interest. The inverse of the
FIM is also used to asymptotically approximate the covariance matrix
of the maximum likelihood estimator. The FIM is also related directly
to the perfomance of the ideal observervon the task of detecting a
small change in the parameter vector of interest, as measured by the
area under the Receiver Operating Characteristic (ROC) curve, also
known as the AUC. The Bayesian version of the FIM is related by the
van Trees inequality to the Ensemble Mean Squared Error (EMSE) for
any estimator of the parameter vector of interest. The lower bound
given by the van Trees inequality is also known as the Bayesian CRB,
and for this reason we will refer to the Bayesian FIM for the relevant
matrix. Elsewhere we have shown that the Bayesian FIM is also directly
related to the average Shannon Information (SI) for the task of detecting
a small change in the parameter vector of interest.

In many cases there are other parameters in the model for the imaging
data that are not of interest for the task at hand. We refer to these
as nuisance parameters and use them to form the components os the
nuisance parameter vector. One way to deal with nuisance parameters
for estimation tasks is to estimate them along with the parameters
of interest and then ignore these estimates. If we have a prior probability
distribution function (PDF) for the nuisance parameter vector, then
we may marginalize over the nuisance parameters to produce a conditional
PDF for the data that only depends on the parameters of interest.
This is the approach that we will examine in this paper to develop
inequalities and approximations for the FIM when the data is affected
by nuisance parameters.

In Section 2 we develop and inequality for the FIM when the prior
PDF on the nuisance parameter vector is independent of the parameters
of interest. In Section 3 we generalize this inequality to the case
where the PDF for the nuisance parameters depends on the parameters
of interest. Section 4 contains description of the implications of
inequalities derived in Section 3 for the Bayesian FIM. Finally in
Section 4 we present an approximation to the FIM that could be useful
if the prior PDF for the nuisance parameters is narrowly distributed
around a nominal value that is known.

\section{Signal independent nuisance parameters}

We will be considering list-mode data in this paper, but everything
is still valid for binned data. For list-mode data the $n^{th}$ detected
photon generates an attribute column vector $\mathbf{a}_{n}$ and
these vectors are then aggregated into a matrix $\mathbf{A}=\left[\mathbf{a}_{1}\cdots\mathbf{a}_{N}\right]$.
An attribute vector may include position of the photon detection,
direction that the photon was travelling when detected, frequancy
and/or polarization state. The number $N$ of photons collected may
be fixed or random. For simplicity we will consider $N$ to be fixed,
but all of the results apply equally well when $N$ is random. We
are interested in either estimating or detecting a small change in
a parameter vector $\boldsymbol{\theta}$ using the data $\mathbf{A}$.
The vector $\boldsymbol{\phi}$ contains nuisance parameters, i.e.,
parameters that affect the data but that we are not interested in
estimating. We assume that we have a prior probability distribution
function (PDF) $pr\left(\boldsymbol{\phi}\right)$ for the nuisance
parameters. This PDF is defined on the $p$-dimensional nuisance parameter
space $\Phi$. The model for the data PDF is then given by
\begin{equation}
pr\left(\mathbf{A}|\boldsymbol{\theta}\right)=\int_{\Phi}pr\left(\mathbf{A}|\boldsymbol{\phi},\boldsymbol{\theta}\right)pr\left(\boldsymbol{\phi}\right)d^{p}\phi
\end{equation}
If the attribute space is partitioned into bins and the number of
photons in each bin is counted then this will produce an integer data
vector $\mathbf{g}$whose dimension is the number of bins. In this
case the data model is given by 
\begin{equation}
Pr\left(\mathbf{g}|\boldsymbol{\theta}\right)=\int_{\Phi}Pr\left(\mathbf{g}|\boldsymbol{\phi},\boldsymbol{\theta}\right)pr\left(\boldsymbol{\phi}\right)d^{p}\phi
\end{equation}
where $Pr\left(\mathbf{g}|\boldsymbol{\theta}\right)$ and $Pr\left(\mathbf{g}|\boldsymbol{\phi},\boldsymbol{\theta}\right)$
are probabilities rather that PDFs. We will be using the list-mode
model throughout, but all of the results will hold with minor notational
changes for binned data.

The score vector $\mathbf{s}\left(\mathbf{A}|\boldsymbol{\theta}\right)$
is defined by $\mathbf{s}\left(\mathbf{A}|\boldsymbol{\theta}\right)=\nabla_{\boldsymbol{\theta}}\ln pr\left(\mathbf{A}|\theta\right)$.
We can write the score vector as
\[
\mathbf{s}\left(\mathbf{A}|\boldsymbol{\theta}\right)=\frac{\int_{\Phi}pr\left(\mathbf{A}|\boldsymbol{\phi},\boldsymbol{\theta}\right)pr\left(\boldsymbol{\phi}\right)\nabla_{\boldsymbol{\theta}}\ln pr\left(\mathbf{A}|\boldsymbol{\phi},\boldsymbol{\theta}\right)d^{p}\phi}{pr\left(\mathbf{A}|\boldsymbol{\theta}\right)}
\]
Using the posterior PDF 
\[
pr\left(\boldsymbol{\phi}|\mathbf{A},\boldsymbol{\theta}\right)=\frac{pr\left(\mathbf{A}|\boldsymbol{\phi},\boldsymbol{\theta}\right)pr\left(\boldsymbol{\phi}\right)}{pr\left(\mathbf{A}|\boldsymbol{\theta}\right)}
\]
and the local score vector
\[
\mathbf{s}\left(\mathbf{A}|\boldsymbol{\phi},\boldsymbol{\theta}\right)=\nabla_{\boldsymbol{\theta}}\ln pr\left(\mathbf{A}|\boldsymbol{\phi},\boldsymbol{\theta}\right)
\]
the (global) score vector is given by
\[
\mathbf{s}\left(\mathbf{A}|\boldsymbol{\theta}\right)=\int_{\Phi}pr\left(\boldsymbol{\phi}|\mathbf{A},\boldsymbol{\theta}\right)\mathbf{s}\left(\mathbf{A}|\boldsymbol{\phi},\boldsymbol{\theta}\right)d^{p}\phi
\]
The Fisher Information Matrix (FIM) is defined by the expectation
{[}1,2{]}
\[
\mathbf{F}\left(\boldsymbol{\theta}\right)=\left\langle \mathbf{s}\left(\mathbf{A}|\boldsymbol{\theta}\right)\mathbf{s}^{\dagger}\left(\mathbf{A}|\boldsymbol{\theta}\right)\right\rangle _{\mathbf{A}|\boldsymbol{\theta}}
\]

This matrix then leads to the well known Cramer-Rao lower bound (CRB)
on the variance of any unbiased estimator of any component of $\boldsymbol{\theta}$.
However, there is another aspect of the FIM which is not as well known.
We have shown previously that for small $\triangle\boldsymbol{\theta}$,
the quantity $\triangle\boldsymbol{\theta}^{\dagger}\mathbf{F}\left(\boldsymbol{\theta}\right)\triangle\boldsymbol{\theta}$
is directly related to the performance of the ideal observer {[}3{]}
(as measured by the area under the ROC curve) on the task of classifying
whether a data matrix $\mathbf{A}$ was generated by the pdf $pr\left(\mathbf{A}|\boldsymbol{\theta}\right)$
or the nearby pdf $pr\left(\mathbf{A}|\boldsymbol{\theta}+\triangle\boldsymbol{\theta}\right)$.
In other words, this scalar quantity measures our ability to detect
small changes in the parameter vector $\boldsymbol{\theta}$ {[}4,5,6{]}.
It is useful to consider this scalar to develop inequalities and then
use them to derive inequalities for the FIM. We now have
\[
\triangle\boldsymbol{\theta}^{\dagger}\mathbf{F}\left(\boldsymbol{\theta}\right)\triangle\boldsymbol{\theta}=\left\langle \left[\int_{\Phi}pr\left(\boldsymbol{\phi}|\mathbf{A},\boldsymbol{\theta}\right)\triangle\boldsymbol{\theta}^{\dagger}\mathbf{s}\left(\mathbf{A}|\boldsymbol{\phi},\boldsymbol{\theta}\right)d^{p}\phi\right]^{2}\right\rangle _{\mathbf{A}|\boldsymbol{\theta}}
\]
This expectation gives us the detectability of a small change in the
parameter of interest. On the other hand, with $\mathbf{s}\left(\mathbf{A}|\boldsymbol{\phi},\boldsymbol{\theta}\right)=\nabla_{\boldsymbol{\theta}}\ln pr\left(\mathbf{A}|\boldsymbol{\phi},\boldsymbol{\theta}\right)$.
the average FIM component for fixed nuisance parameters is 
\[
\left\langle \triangle\boldsymbol{\theta}^{\dagger}\mathbf{F}\left(\boldsymbol{\phi},\boldsymbol{\theta}\right)\triangle\boldsymbol{\theta}\right\rangle _{\boldsymbol{\phi}}=\int_{\Phi}pr\left(\boldsymbol{\phi}\right)\left\langle \left[\triangle\boldsymbol{\theta}^{\dagger}\mathbf{s}\left(\mathbf{A}|\boldsymbol{\phi},\boldsymbol{\theta}\right)\right]^{2}\right\rangle _{\mathbf{A}|\boldsymbol{\phi},\boldsymbol{\theta}}d^{p}\phi
\]
We can write this as
\[
\left\langle \triangle\boldsymbol{\theta}^{\dagger}\mathbf{F}\left(\boldsymbol{\phi},\boldsymbol{\theta}\right)\triangle\boldsymbol{\theta}\right\rangle _{\boldsymbol{\phi}}=\left\langle \int_{\Phi}pr\left(\boldsymbol{\phi}|\mathbf{A},\boldsymbol{\theta}\right)\left[\triangle\boldsymbol{\theta}^{\dagger}\mathbf{s}\left(\mathbf{A}|\boldsymbol{\phi},\boldsymbol{\theta}\right)\right]^{2}d^{p}\phi\right\rangle _{\mathbf{A}|\boldsymbol{\theta}}
\]
This would apply to the case where the nuisance parameters vary randomly,
but we have some way to measure them in each case. This expectation
gives us the average detectability of a small change in the parameter
of interest, where the average is over the random nuisance parameters.
The difference between these two quantities is positive and given
by
\[
\left\langle \triangle\boldsymbol{\theta}^{\dagger}\mathbf{F}\left(\boldsymbol{\phi},\boldsymbol{\theta}\right)\triangle\boldsymbol{\theta}\right\rangle _{\boldsymbol{\boldsymbol{\phi}}}-\triangle\boldsymbol{\theta}^{\dagger}\mathbf{F}\left(\boldsymbol{\theta}\right)\triangle\boldsymbol{\theta}=\left\langle var\left[\triangle\boldsymbol{\theta}^{\dagger}\mathbf{s}\left(\mathbf{A}|\boldsymbol{\phi},\boldsymbol{\theta}\right)|\mathbf{A},\boldsymbol{\theta}\right]\right\rangle _{\mathbf{A}|\boldsymbol{\theta}}
\]
This gives us the bound
\[
\triangle\boldsymbol{\theta}^{\dagger}\mathbf{F}\left(\boldsymbol{\theta}\right)\triangle\boldsymbol{\theta}\leq\left\langle \triangle\boldsymbol{\theta}^{\dagger}\mathbf{F}\left(\boldsymbol{\phi},\boldsymbol{\theta}\right)\triangle\boldsymbol{\theta}\right\rangle _{\boldsymbol{\phi}}
\]
This is not an unexpected result. It says that, on average, it is
better to know the values of the nuisance parameters when we are trying
to detect a change in the parameters of interest. In terms of the
FIMs themselves we have the relation
\[
\mathbf{\left\langle \mathbf{F}\left(\boldsymbol{\phi},\boldsymbol{\theta}\right)\right\rangle _{\boldsymbol{\boldsymbol{\phi}}}-F}\left(\boldsymbol{\theta}\right)=\left\langle cov\left[\mathbf{s}\left(\mathbf{A}|\boldsymbol{\phi},\boldsymbol{\theta}\right)|\mathbf{A},\boldsymbol{\theta}\right]\right\rangle _{\mathbf{A}|\boldsymbol{\theta}}
\]
and therefore the matrix inequality
\[
\mathbf{F}\left(\boldsymbol{\theta}\right)\leq\left\langle \mathbf{F}\left(\boldsymbol{\phi},\boldsymbol{\theta}\right)\right\rangle _{\boldsymbol{\boldsymbol{\phi}}}
\]
For the Cramer-Rao bound we then have
\[
\left[\mathbf{F}\left(\boldsymbol{\theta}\right)\right]^{-1}\geq\left[\left\langle \mathbf{F}\left(\boldsymbol{\phi},\boldsymbol{\theta}\right)\right\rangle _{\boldsymbol{\boldsymbol{\phi}}}\right]^{-1}
\]
It is not as easy to provide an interpretation for this inequality
since, on the right, we are averaging over the nuisance parameters
before we take the inverse. We would like to say that the average
CRB when we know the nuisance parameters is less than the CRB when
we do not, but this statement would require averaging over the nuisance
parameters after we take the inverse of the matrices $\mathbf{F}\left(\boldsymbol{\phi},\boldsymbol{\theta}\right)$.

\section{Signal dependent nuisance parameters}

We may be faced with a situation where the PDF for the nuisance parameters
depends on the parameters of interest. In this case we have

\[
pr\left(\mathbf{A}|\boldsymbol{\theta}\right)=\int_{\Phi}pr\left(\mathbf{A}|\boldsymbol{\phi},\boldsymbol{\theta}\right)pr\left(\boldsymbol{\phi}|\boldsymbol{\theta}\right)d^{p}\phi
\]
The score vector is given by
\[
\mathbf{s}\left(\mathbf{A}|\boldsymbol{\theta}\right)=\frac{\int_{\Phi}pr\left(\mathbf{A}|\boldsymbol{\phi},\boldsymbol{\theta}\right)pr\left(\boldsymbol{\phi}|\boldsymbol{\theta}\right)\nabla_{\boldsymbol{\theta}}\ln pr\left(\mathbf{A}|\boldsymbol{\phi},\boldsymbol{\theta}\right)d^{p}\phi}{pr\left(\mathbf{A}|\boldsymbol{\theta}\right)}+\frac{\int_{\Phi}pr\left(\mathbf{A}|\boldsymbol{\phi},\boldsymbol{\theta}\right)pr\left(\boldsymbol{\phi}|\boldsymbol{\theta}\right)\nabla_{\boldsymbol{\theta}}\ln pr\left(\boldsymbol{\phi}|\boldsymbol{\theta}\right)d^{p}\phi}{pr\left(\mathbf{A}|\boldsymbol{\theta}\right)}
\]
We write this as
\[
\mathbf{s}\left(\mathbf{A}|\boldsymbol{\theta}\right)=\int_{\Phi}pr\left(\boldsymbol{\phi}|\mathbf{A},\boldsymbol{\theta}\right)\mathbf{s}\left(\mathbf{A}|\boldsymbol{\phi},\boldsymbol{\theta}\right)d^{p}\phi+\int_{\Phi}pr\left(\boldsymbol{\phi}|\mathbf{A},\boldsymbol{\theta}\right)\mathbf{s}\left(\boldsymbol{\phi}|\boldsymbol{\theta}\right)d^{p}\phi
\]
with 
\[
\mathbf{s}\left(\mathbf{A}|\boldsymbol{\phi},\boldsymbol{\theta}\right)=\nabla_{\boldsymbol{\theta}}\ln pr\left(\mathbf{A}|\boldsymbol{\phi},\boldsymbol{\theta}\right)
\]
and
\[
\mathbf{s}\left(\boldsymbol{\phi}|\boldsymbol{\theta}\right)=\nabla_{\boldsymbol{\theta}}\ln pr\left(\boldsymbol{\phi}|\boldsymbol{\theta}\right)
\]
We now have
\[
\triangle\boldsymbol{\theta}^{\dagger}\mathbf{F}\left(\boldsymbol{\theta}\right)\triangle\boldsymbol{\theta}=\left\langle \left\langle \triangle\boldsymbol{\theta}^{\dagger}\mathbf{s}\left(\mathbf{A}|\boldsymbol{\phi},\boldsymbol{\theta}\right)+\triangle\boldsymbol{\theta}^{\dagger}\mathbf{s}\left(\boldsymbol{\phi}|\boldsymbol{\theta}\right)\right\rangle _{\boldsymbol{\phi}|\mathbf{A},\boldsymbol{\theta}}^{2}\right\rangle _{\mathbf{A}|\boldsymbol{\theta}}
\]
Since the variance of any random variable is always nonegative we
can write
\[
\triangle\boldsymbol{\theta}^{\dagger}\mathbf{F}\left(\boldsymbol{\theta}\right)\triangle\boldsymbol{\theta}\leq\left\langle \left\langle \left[\triangle\boldsymbol{\theta}^{\dagger}\mathbf{s}\left(\mathbf{A}|\boldsymbol{\phi},\boldsymbol{\theta}\right)+\triangle\boldsymbol{\theta}^{\dagger}\mathbf{s}\left(\boldsymbol{\phi}|\boldsymbol{\theta}\right)\right]^{2}\right\rangle _{\boldsymbol{\phi}|\mathbf{A},\boldsymbol{\theta}}\right\rangle _{\mathbf{A}|\boldsymbol{\theta}}
\]
Now we reverse the order of the expectations and have
\[
\triangle\boldsymbol{\theta}^{\dagger}\mathbf{F}\left(\boldsymbol{\theta}\right)\triangle\boldsymbol{\theta}\leq\left\langle \left\langle \left[\triangle\boldsymbol{\theta}^{\dagger}\mathbf{s}\left(\mathbf{A}|\boldsymbol{\phi},\boldsymbol{\theta}\right)+\triangle\boldsymbol{\theta}^{\dagger}\mathbf{s}\left(\boldsymbol{\phi}|\boldsymbol{\theta}\right)\right]^{2}\right\rangle _{\mathbf{A}|\boldsymbol{\phi},\boldsymbol{\theta}}\right\rangle _{\boldsymbol{\phi}|\boldsymbol{\theta}}
\]
Since the mean of the score vector is zero, the cross term vanishes
when we expand the square of the term in square brackets. Thus we
have
\[
\triangle\boldsymbol{\theta}^{\dagger}\mathbf{F}\left(\boldsymbol{\theta}\right)\triangle\boldsymbol{\theta}\leq\left\langle \triangle\boldsymbol{\theta}^{\dagger}\mathbf{F}\left(\boldsymbol{\phi},\boldsymbol{\theta}\right)\triangle\boldsymbol{\theta}\right\rangle _{\boldsymbol{\phi}}+\left\langle \left[\triangle\boldsymbol{\theta}^{\dagger}\mathbf{s}\left(\boldsymbol{\phi}|\boldsymbol{\theta}\right)\right]^{2}\right\rangle _{\boldsymbol{\phi}|\boldsymbol{\theta}}
\]
On the right side we have two contributions. One would give us the
average over the nuisance parameters of the square of the detectability
of a small change in the parameter of interest if we knew the random
nuisance parameters via some other measurement. The second term gives
us the square of the detectability of a small change in the parameters
of interest from the measurement of the nuisance parameters themselves.
From the first inequality the difference between the right-hand side
and lieft-hand side in the second inequality is given by the average
variance:
\[
\left\langle var_{\boldsymbol{\phi}}\left[\triangle\boldsymbol{\theta}^{\dagger}\mathbf{s}\left(\mathbf{A}|\boldsymbol{\phi},\boldsymbol{\theta}\right)+\triangle\boldsymbol{\theta}^{\dagger}\mathbf{s}\left(\boldsymbol{\phi}|\boldsymbol{\theta}\right)|\mathbf{A},\boldsymbol{\theta}\right]\right\rangle _{\mathbf{A}|\boldsymbol{\theta}}
\]
The subscript on the variance function indicates the random vector
whose PDF is used in the expectations to compute that variance. The
vectors to the right of the vertical bar in the variance function
indicate what vectors are being held fixed when computing the expectations.
In terms of the relevant matrices we now can write

\[
\mathbf{F}\left(\boldsymbol{\theta}\right)\leq\left\langle \mathbf{F}\left(\boldsymbol{\phi},\boldsymbol{\theta}\right)\right\rangle _{\boldsymbol{\boldsymbol{\phi}}|\boldsymbol{\theta}}+\left\langle \left[\nabla_{\boldsymbol{\theta}}\ln pr\left(\boldsymbol{\phi}|\boldsymbol{\theta}\right)\right]\left[\nabla_{\boldsymbol{\theta}}\ln pr\left(\boldsymbol{\phi}|\boldsymbol{\theta}\right)\right]^{\dagger}\right\rangle _{\boldsymbol{\boldsymbol{\phi}}|\boldsymbol{\theta}}
\]
The last term on the right is also a Fisher information matrix $\mathbf{F}_{\boldsymbol{\phi}}\left(\boldsymbol{\theta}\right)$.
This matrix measures the information contained in the nuisance parameters
about the parameters of interest. Now we can write this inequality
as
\[
\mathbf{F}\left(\boldsymbol{\theta}\right)\leq\left\langle \mathbf{F}\left(\boldsymbol{\phi},\boldsymbol{\theta}\right)\right\rangle _{\boldsymbol{\boldsymbol{\phi}}|\boldsymbol{\theta}}+\mathbf{F}_{\boldsymbol{\phi}}\left(\boldsymbol{\theta}\right)
\]
 It is difficult to interpret this inequality in terms of the CRB
since the inverse of the sum of two matrices is not easily relatable
to the sum of their inverses. The difference between the matrix on
the right and the one on the left in this inequality is a covariance
matrix
\[
\left\langle cov_{\boldsymbol{\phi}}\left[\mathbf{s}\left(\mathbf{A}|\boldsymbol{\phi},\boldsymbol{\theta}\right)+\mathbf{s}\left(\boldsymbol{\phi}|\boldsymbol{\theta}\right)|\mathbf{A},\boldsymbol{\theta}\right]\right\rangle _{\mathbf{A}|\boldsymbol{\theta}}
\]

\section{Relation to Bayesian FIM for joint model}

The averages of FIMs that appear in the previous two sections are
reminiscent of averages over parameters that appear in the van Trees
inequality, also known as the Bayesian CRB {[}7,8{]}. For this reason
we will call the version of the FIM that appears in the Bayesian CRB
the Bayesian FIM. To compute the Bayesian FIM we need a prior distribution
$pr\left(\boldsymbol{\theta}\right)$ on the parameters of interest.
The van Trees inequality then uses the Bayesian FIM to provide a lower
bound for the EMSE when we are estimating these parameters. This inequality,
like the CRB, requires inverting the Bayesian FIM. We can also show
that the Bayesian FIM, without inversion, is directly related to the
average Shannon information for the task of detecting a small change
in the parameters of interest. For the Bayesian FIM we define posterior
score vectors via the posterior distribution as $\mathbf{s}_{\boldsymbol{\theta}}=\nabla_{\boldsymbol{\theta}}\ln pr\left(\boldsymbol{\theta},\boldsymbol{\phi}|\mathbf{A}\right)$
and $\mathbf{s}_{\boldsymbol{\phi}}=\nabla_{\boldsymbol{\phi}}\ln pr\left(\boldsymbol{\theta},\boldsymbol{\phi}|\mathbf{A}\right)$.
The Bayesian FIM for the pair $\left(\boldsymbol{\theta},\boldsymbol{\phi}\right)$
is then given by the matrix

\[
\mathbf{F}_{J}=\left\langle \left\langle \left\langle \left[\begin{array}{cc}
\mathbf{s}_{\boldsymbol{\theta}}\mathbf{s}_{\boldsymbol{\theta}}^{\dagger} & \mathbf{s}_{\boldsymbol{\theta}}\mathbf{s}_{\boldsymbol{\phi}}^{\dagger}\\
\mathbf{s}_{\boldsymbol{\phi}}\mathbf{s}_{\boldsymbol{\theta}}^{\dagger} & \mathbf{s}_{\boldsymbol{\phi}}\mathbf{s}_{\boldsymbol{\phi}}^{\dagger}
\end{array}\right]\right\rangle _{\mathbf{A}|\boldsymbol{\phi},\boldsymbol{\theta}}\right\rangle _{\boldsymbol{\phi}|\boldsymbol{\theta}}\right\rangle _{\boldsymbol{\theta}}=\left[\begin{array}{cc}
\mathbf{F}_{\boldsymbol{\theta\theta}} & \mathbf{F}_{\boldsymbol{\theta\phi}}\\
\mathbf{F}_{\boldsymbol{\phi\theta}} & \mathbf{F}_{\boldsymbol{\phi\phi}}
\end{array}\right]
\]
The subscript $J$ here refers to the fact that the corresponding
estimation problem in the van Trees inequality would be estimating
the pair $\left(\boldsymbol{\theta},\boldsymbol{\phi}\right)$ jointly.
This corresponds to one approach to estimating $\boldsymbol{\theta}$
in the presence of nuisance parameters contained in $\boldsymbol{\phi}$,
which is to estimate both vectors and then ignore the estimate of
the nuisance vector.

Using the definition of the posterior distribution
\[
pr\left(\boldsymbol{\theta},\boldsymbol{\phi}|\mathbf{A}\right)=\frac{pr\left(\mathbf{A}|\boldsymbol{\phi},\boldsymbol{\theta}\right)pr\left(\boldsymbol{\phi}|\boldsymbol{\theta}\right)pr\left(\boldsymbol{\theta}\right)}{pr\left(\mathbf{A}\right)}
\]
we find that the posterior score vectors are given by
\[
\mathbf{s}_{\boldsymbol{\theta}}=\frac{\nabla_{\theta}pr\left(\mathbf{A}|\boldsymbol{\phi},\boldsymbol{\theta}\right)}{pr\left(\mathbf{A}|\boldsymbol{\phi},\boldsymbol{\theta}\right)}+\frac{\nabla_{\theta}pr\left(\boldsymbol{\phi}|\boldsymbol{\theta}\right)}{pr\left(\boldsymbol{\phi}|\boldsymbol{\theta}\right)}+\frac{\nabla_{\theta}pr\left(\boldsymbol{\theta}\right)}{pr\left(\boldsymbol{\theta}\right)}
\]
and 
\[
\mathbf{s}_{\boldsymbol{\phi}}=\frac{\nabla_{\phi}pr\left(\mathbf{A}|\boldsymbol{\phi},\boldsymbol{\theta}\right)}{pr\left(\mathbf{A}|\boldsymbol{\phi},\boldsymbol{\theta}\right)}+\frac{\nabla_{\phi}pr\left(\boldsymbol{\phi}|\boldsymbol{\theta}\right)}{pr\left(\boldsymbol{\phi}|\boldsymbol{\theta}\right)}.
\]
The matrix $\mathbf{F}_{\boldsymbol{\theta}\boldsymbol{\theta}}$
is a sum of three components: $\mathbf{F}_{\boldsymbol{\theta}\boldsymbol{\theta}}=\left\langle \left\langle \mathbf{F}_{11}\left(\boldsymbol{\theta},\boldsymbol{\phi}\right)\right\rangle _{\boldsymbol{\phi}|\boldsymbol{\theta}}\right\rangle _{\boldsymbol{\theta}}+\left\langle \mathbf{F}_{11}\left(\boldsymbol{\theta}\right)\right\rangle _{\boldsymbol{\theta}}+\mathbf{F}_{11}.$
The three matrices appearing on the right in this equation are
\[
\mathbf{F}_{11}\left(\boldsymbol{\theta},\boldsymbol{\phi}\right)=\left\langle \left[\frac{\nabla_{\theta}pr\left(\mathbf{A}|\boldsymbol{\phi},\boldsymbol{\theta}\right)}{pr\left(\mathbf{A}|\boldsymbol{\phi},\boldsymbol{\theta}\right)}\right]\left[\frac{\nabla_{\theta}pr\left(\mathbf{A}|\boldsymbol{\phi},\boldsymbol{\theta}\right)}{pr\left(\mathbf{A}|\boldsymbol{\phi},\boldsymbol{\theta}\right)}\right]^{\dagger}\right\rangle _{\mathbf{A}|\boldsymbol{\phi},\boldsymbol{\theta}},
\]
\[
\mathbf{F}_{11}\left(\boldsymbol{\theta}\right)=\left\langle \left[\frac{\nabla_{\theta}pr\left(\boldsymbol{\phi}|\boldsymbol{\theta}\right)}{pr\left(\boldsymbol{\phi}|\boldsymbol{\theta}\right)}\right]\left[\frac{\nabla_{\theta}pr\left(\boldsymbol{\phi}|\boldsymbol{\theta}\right)}{pr\left(\boldsymbol{\phi}|\boldsymbol{\theta}\right)}\right]^{\dagger}\right\rangle _{\boldsymbol{\phi}|\boldsymbol{\theta}}
\]
and
\[
\mathbf{F}_{11}=\left\langle \left[\frac{\nabla_{\boldsymbol{\theta}}pr\left(\boldsymbol{\theta}\right)}{pr\left(\boldsymbol{\theta}\right)}\right]\left[\frac{\nabla_{\boldsymbol{\theta}}pr\left(\boldsymbol{\theta}\right)}{pr\left(\boldsymbol{\theta}\right)}\right]^{\dagger}\right\rangle _{\boldsymbol{\theta}}.
\]
This is the principal matrix of interest, but for completeness we
also provide: $\mathbf{F}_{\phi\phi}=\left\langle \left\langle \mathbf{F}_{22}\left(\boldsymbol{\theta},\boldsymbol{\phi}\right)\right\rangle _{\boldsymbol{\phi}|\boldsymbol{\theta}}\right\rangle _{\boldsymbol{\theta}}+\left\langle \mathbf{F}_{22}\left(\boldsymbol{\theta}\right)\right\rangle _{\boldsymbol{\theta}}$
with
\[
\mathbf{F}_{22}\left(\boldsymbol{\theta},\boldsymbol{\phi}\right)=\left\langle \left[\frac{\nabla_{\phi}pr\left(\mathbf{A}|\boldsymbol{\phi},\boldsymbol{\theta}\right)}{pr\left(\mathbf{A}|\boldsymbol{\phi},\boldsymbol{\theta}\right)}\right]\left[\frac{\nabla_{\phi}pr\left(\mathbf{A}|\boldsymbol{\phi},\boldsymbol{\theta}\right)}{pr\left(\mathbf{A}|\boldsymbol{\phi},\boldsymbol{\theta}\right)}\right]^{\dagger}\right\rangle _{\mathbf{A}|\boldsymbol{\phi},\boldsymbol{\theta}}
\]
and
\[
\mathbf{F}_{22}\left(\boldsymbol{\theta}\right)=\left\langle \left[\frac{\nabla_{\phi}pr\left(\boldsymbol{\phi}|\boldsymbol{\theta}\right)}{pr\left(\boldsymbol{\phi}|\boldsymbol{\theta}\right)}\right]\left[\frac{\nabla_{\phi}pr\left(\boldsymbol{\phi}|\boldsymbol{\theta}\right)}{pr\left(\boldsymbol{\phi}|\boldsymbol{\theta}\right)}\right]^{\dagger}\right\rangle _{\boldsymbol{\phi}|\boldsymbol{\theta}}.
\]
We also have $\mathbf{F}_{\boldsymbol{\theta}\boldsymbol{\phi}}=\left\langle \left\langle \mathbf{F}_{12}\left(\boldsymbol{\theta},\boldsymbol{\phi}\right)\right\rangle _{\boldsymbol{\phi}|\boldsymbol{\theta}}\right\rangle _{\boldsymbol{\theta}}+\left\langle \mathbf{F}_{12}\left(\boldsymbol{\theta}\right)\right\rangle _{\boldsymbol{\theta}}$
with
\[
\mathbf{F}_{12}\left(\boldsymbol{\theta},\boldsymbol{\phi}\right)=\left\langle \left[\frac{\nabla_{\theta}pr\left(\mathbf{A}|\boldsymbol{\phi},\boldsymbol{\theta}\right)}{pr\left(\mathbf{A}|\boldsymbol{\phi},\boldsymbol{\theta}\right)}\right]\left[\frac{\nabla_{\phi}pr\left(\mathbf{A}|\boldsymbol{\phi},\boldsymbol{\theta}\right)}{pr\left(\mathbf{A}|\boldsymbol{\phi},\boldsymbol{\theta}\right)}\right]^{\dagger}\right\rangle _{\mathbf{A}|\boldsymbol{\phi},\boldsymbol{\theta}}
\]
and
\[
\mathbf{F}_{12}\left(\boldsymbol{\theta}\right)=\left\langle \left[\frac{\nabla_{\theta}pr\left(\boldsymbol{\phi}|\boldsymbol{\theta}\right)}{pr\left(\boldsymbol{\phi}|\boldsymbol{\theta}\right)}\right]\left[\frac{\nabla_{\phi}pr\left(\boldsymbol{\phi}|\boldsymbol{\theta}\right)}{pr\left(\boldsymbol{\phi}|\boldsymbol{\theta}\right)}\right]^{\dagger}\right\rangle _{\boldsymbol{\phi}|\boldsymbol{\theta}}.
\]
The component $\mathbf{F}_{\boldsymbol{\phi}\boldsymbol{\theta}}$
is the transpose of $\mathbf{F}_{\boldsymbol{\theta}\boldsymbol{\phi}}$. 

From the results in Section 3 we have, in the notation in this section,
\[
\mathbf{F}\left(\boldsymbol{\theta}\right)\leq\left\langle \mathbf{F}_{11}\left(\boldsymbol{\theta},\boldsymbol{\phi}\right)\right\rangle _{\boldsymbol{\phi}|\boldsymbol{\theta}}+\mathbf{F}_{11}\left(\boldsymbol{\theta}\right).
\]
The Bayesian FIM for the model in Section 3, where we are marginalizing
over $\boldsymbol{\phi}$, is defined by
\[
\mathbf{F}_{M}=\left\langle \left\langle \left[\nabla_{\boldsymbol{\theta}}\ln pr\left(\boldsymbol{\theta}|\mathbf{A}\right)\right]\left[\nabla_{\boldsymbol{\theta}}\ln pr\left(\boldsymbol{\theta}|\mathbf{A}\right)\right]^{\dagger}\right\rangle _{\mathbf{A}|\boldsymbol{\theta}}\right\rangle _{\boldsymbol{\theta}}.
\]
The subscript $M$ refers to the marginalization of the nuisance parameters
in the model. We now have $\mathbf{F}_{M}=\left\langle \mathbf{F}\left(\boldsymbol{\theta}\right)\right\rangle _{\boldsymbol{\theta}}+\mathbf{F}_{11}$,
which implies that $\mathbf{F}_{M}\leq\mathbf{F}_{\boldsymbol{\theta}\boldsymbol{\theta}}$.
If we are trying to detect a small change $\triangle\boldsymbol{\theta}$
in the parameter vector of interest, then we have shown that $\triangle\boldsymbol{\theta}^{\dagger}\mathbf{F}_{M}\triangle\boldsymbol{\theta}$
is a useful figure of merit related to the average SI {[}9{]} for
this detection task {[}10{]}. This SI can in turn be related to the
ideal observer AUC via an integral transform {[}11,12{]}. We then
have 
\[
\triangle\boldsymbol{\theta}^{\dagger}\mathbf{F}_{M}\triangle\boldsymbol{\theta}\leq\triangle\boldsymbol{\theta}^{\dagger}\mathbf{F}_{\boldsymbol{\theta}\boldsymbol{\theta}}\triangle\boldsymbol{\theta}=\left[\begin{array}{c}
\triangle\boldsymbol{\theta}\\
\mathrm{0}
\end{array}\right]^{\dagger}\mathbf{F}_{J}\left[\begin{array}{c}
\triangle\boldsymbol{\theta}\\
\mathrm{0}
\end{array}\right]
\]
On the right in this inequality is the same figure of merit when $\boldsymbol{\phi}$
is not marginalized out and $\triangle\boldsymbol{\mathbf{\phi}}=0.$
We have therefore the not too surprising conclusion that our ability
to detect a change in the parameter vector of interest is increased,
on average, if we know the value of the random nuisance parameter
vector.

\section{Approximate change in the FIM due to nuisance parameter uncertainty}

There are situations where we have nominal values for the nuisance
parameters but there is still some uncertainty in their actual values.
If the nominal values are the components of the vector $\boldsymbol{\phi}$,
then the probability of the data conditional on the parameters of
interest is given by 
\[
pr\left(\mathbf{A}|\boldsymbol{\theta}\right)=\int_{\Omega}pr\left(\mathbf{A}|\boldsymbol{\phi}+\triangle\boldsymbol{\phi},\boldsymbol{\theta}\right)pr\left(\triangle\boldsymbol{\phi}\right)d^{q}\triangle\phi
\]
We will assume that the mean of the error vector $\triangle\boldsymbol{\phi}$
is zero. We want to find an approximation to the FIM that will be
useful if the PDF $pr\left(\triangle\boldsymbol{\phi}\right)$ is
concentrated around the origin in the nuisance parameter space.

We start with Taylor series expansion for the PDF $pr\left(\mathbf{A}|\boldsymbol{\theta}\right)$:
\[
pr\left(\mathbf{A}|\boldsymbol{\theta}\right)=\int_{\Omega}\left[pr\left(\mathbf{A}|\boldsymbol{\phi},\boldsymbol{\theta}\right)+\triangle\boldsymbol{\phi}^{\dagger}\nabla_{\boldsymbol{\phi}}pr\left(\mathbf{A}|\boldsymbol{\phi},\boldsymbol{\theta}\right)+\frac{1}{2}\triangle\boldsymbol{\phi}^{\dagger}\nabla_{\boldsymbol{\phi}}\nabla_{\boldsymbol{\phi}}^{\dagger}pr\left(\mathbf{A}|\boldsymbol{\phi},\boldsymbol{\theta}\right)\triangle\boldsymbol{\phi}+\ldots\right]pr\left(\triangle\boldsymbol{\phi}\right)d^{q}\triangle\phi
\]
Using the fact that the mean error vector is zero we find the lowest
order terms;
\[
pr\left(\mathbf{A}|\boldsymbol{\theta}\right)=pr\left(\mathbf{A}|\boldsymbol{\phi},\boldsymbol{\theta}\right)+\frac{1}{2}\mathrm{tr}\left[\mathbf{K}_{\boldsymbol{\phi}}\nabla_{\boldsymbol{\phi}}\nabla_{\boldsymbol{\phi}}^{\dagger}pr\left(\mathbf{A}|\boldsymbol{\phi},\boldsymbol{\theta}\right)\right]+\ldots
\]
In this equation the matrix $\mathbf{K}_{\boldsymbol{\phi}}$ is the
covariance matrix for $\triangle\boldsymbol{\phi}$ . We may now derive
an approximate expression for the FIM which makes use of the constant
coefficient second order differential operator
\[
\mathcal{L}_{\boldsymbol{\phi}}=\frac{1}{2}\nabla_{\boldsymbol{\phi}}^{\dagger}\mathbf{K}_{\boldsymbol{\phi}}\nabla_{\boldsymbol{\phi}}.
\]
We can write to lowest order in $\mathbf{K}_{\boldsymbol{\phi}}$,
\[
pr\left(\mathbf{A}|\boldsymbol{\theta}\right)=pr\left(\mathbf{A}|\boldsymbol{\phi},\boldsymbol{\theta}\right)+\mathcal{L}_{\boldsymbol{\phi}}pr\left(\mathbf{A}|\boldsymbol{\phi},\boldsymbol{\theta}\right)+\ldots
\]
Now we can use the series expansion for the logarithm to write
\[
\ln pr\left(\mathbf{A}|\boldsymbol{\theta}\right)=\ln pr\left(\mathbf{A}|\boldsymbol{\phi},\boldsymbol{\theta}\right)+\frac{\mathcal{L}_{\boldsymbol{\phi}}pr\left(\mathbf{A}|\boldsymbol{\phi},\boldsymbol{\theta}\right)}{pr\left(\mathbf{A}|\boldsymbol{\phi},\boldsymbol{\theta}\right)}+\ldots
\]
We will use these two expansions to derive a series expansion for
the FIM $\mathbf{F}\left(\boldsymbol{\theta}\right)$.

By definition the FIM in question is given by
\[
\mathbf{F}\left(\boldsymbol{\theta}\right)=\int_{D}\left[\nabla_{\boldsymbol{\theta}}\ln pr\left(\mathbf{A}|\boldsymbol{\theta}\right)\right]\left[\nabla_{\boldsymbol{\theta}}\ln pr\left(\mathbf{A}|\boldsymbol{\theta}\right)\right]^{\dagger}pr\left(\mathbf{A}|\boldsymbol{\theta}\right)d^{M}g
\]
To lowest order there are three correction terms

\[
\mathbf{F}\left(\boldsymbol{\theta}\right)=\mathbf{F}\left(\boldsymbol{\phi},\boldsymbol{\theta}\right)+\mathbf{F}_{1}\left(\boldsymbol{\phi},\boldsymbol{\theta}\right)+\mathbf{F}_{2}\left(\boldsymbol{\phi},\boldsymbol{\theta}\right)-\mathbf{F}_{3}\left(\boldsymbol{\phi},\boldsymbol{\theta}\right)+\ldots
\]
The correction terms in this expansion are
\[
\mathbf{F}_{1}\left(\boldsymbol{\phi},\boldsymbol{\theta}\right)=\int_{D}\left[\nabla_{\boldsymbol{\theta}}\ln pr\left(\mathbf{A}|\boldsymbol{\phi},\boldsymbol{\theta}\right)\right]\left[\nabla_{\boldsymbol{\theta}}\mathcal{L}_{\boldsymbol{\phi}}pr\left(\mathbf{A}|\boldsymbol{\phi},\boldsymbol{\theta}\right)\right]^{\dagger}d^{M}g,
\]
\[
\mathbf{F}_{2}\left(\boldsymbol{\phi},\boldsymbol{\theta}\right)=\int_{D}\left[\nabla_{\boldsymbol{\theta}}\mathcal{L}_{\boldsymbol{\phi}}pr\left(\mathbf{A}|\boldsymbol{\phi},\boldsymbol{\theta}\right)\right]\left[\nabla_{\boldsymbol{\theta}}\ln pr\left(\mathbf{A}|\boldsymbol{\phi},\boldsymbol{\theta}\right)\right]^{\dagger}d^{M}g
\]
and
\[
\mathbf{F}_{3}\left(\boldsymbol{\phi},\boldsymbol{\theta}\right)=\int_{D}\left[\nabla_{\boldsymbol{\theta}}\ln pr\left(\mathbf{A}|\boldsymbol{\phi},\boldsymbol{\theta}\right)\right]\left[\nabla_{\boldsymbol{\theta}}\ln pr\left(\mathbf{A}|\boldsymbol{\phi},\boldsymbol{\theta}\right)\right]^{\dagger}\mathcal{L}_{\boldsymbol{\phi}}pr\left(\mathbf{A}|\boldsymbol{\phi},\boldsymbol{\theta}\right)d^{M}g.
\]
These correction terms can be computed numerically using the same
Monte Carlo methods commonly used to compute the FIM $\mathbf{F}\left(\boldsymbol{\phi},\boldsymbol{\theta}\right)$
. 

The expansion derived here and the inequalities derived above leave
open the possibility that $\mathbf{F}\left(\boldsymbol{\theta}\right)>\mathbf{F}\left(\boldsymbol{\phi},\boldsymbol{\theta}\right)$
for some particular value $\boldsymbol{\phi}$ of the nuisance parameter.
Does this inequality make sense? To see that it can be a valid inequality
consider a particular estimation task. Suppose that we have an imaging
system at one end of an L-shaped hallway and that there is a small
light source around the corner of that hallway. We want to know the
location of that light source, so this is the parameter vector of
interest $\boldsymbol{\theta}$. Assume that there is a swinging door
in the leg of the hallway occupied by the source and that this door,
if closed, completely blocks any light from the source from reaching
our imaging system. The nuisance parameter $\phi$ will be the angle
that the door makes with the wall it is attached to, so that $\phi=0$
or $\phi=\pi$ when the door is wide open, and $\phi=\pi/2$ when
it is closed. If the nominal value for the nuisnace parameter is $\phi=\pi/2$,
then $\mathbf{F}\left(\boldsymbol{\phi},\boldsymbol{\theta}\right)=\mathbf{0}$.
In other words, if we are certain that the door is closed, then the
FIM for $\boldsymbol{\theta}$ is the zero matrix, since no light
from the source is reaching our detector. On the other hand if there
is some uncertainty in $\phi$, i.e. if the door might be open a little
by some random angle $\triangle\phi$, then the FIM $\mathbf{F}\left(\boldsymbol{\theta}\right)$
is not the zero matrix, since the PDF $pr\left(\mathbf{A}|\boldsymbol{\theta}\right)$
includes contributions from configurations where the door is open
a little. This does not violate the inequality $\mathbf{F}\left(\boldsymbol{\theta}\right)\leq\left\langle \mathbf{F}\left(\boldsymbol{\phi},\boldsymbol{\theta}\right)\right\rangle _{\boldsymbol{\boldsymbol{\phi}}}$
in Section 2 since the matrix on the right also includes contributions
from configurations where the door is open a little. 

\section{Conclusion}

In Section 2 we developed and inequality for the FIM when the prior
PDF on the nuisance parameter vector is independent of the parameters
of interest. This inequality states the the FIM for estimating $\boldsymbol{\theta}$,
the vector parametr of interest is nsmaller then the average FIM for
the pair $\left(\boldsymbol{\theta},\boldsymbol{\phi}\right)$, where
$\left(\boldsymbol{\phi}\right)$ is the vector of nuisance parameters
and the average is over these nuisance parameters. Section 3 we generalized
this inequality to the case where the PDF for the nuisance parameters
depends on the parameters of interest. In this case the inequality
involves an extra term which is the FIM for the conditional PDF $pr\left(\boldsymbol{\phi}|\boldsymbol{\theta}\right)$.
In section 4 we described the implications of inequalities derived
in Section 3 for the Bayesian FIM. we found, not surprisingly, that
in terms of the Bayesian FIM it is, on average, better if we know
the values of the nuisance parameters than if we do not. This is,
however, an avergae result which can be violated for individual instances
of the nuisance parameter vector. Finally, in Section 4 we presented
an approximation to the FIM that could be useful if the prior PDF
for the nuisance parameters is narrowly distributed around a nominal
value that is known. This approximation only requires knowledge of
the covariance matrix of the distribution of the nuisance parameter
vecotr around its mean, which is assumed to be the nominal value. 

The inequalities provide upper bounds for the FIM and Bayesian FIM
when nuisance parameters are present. In general, these upper bounds
are easier to compute than the FIM in question. When the nuisance
parameters are known to within some error, then the approximation
could be useful and is again easier to compute than the original FIM.

\end{document}